\documentclass{article}

\usepackage[utf8]{inputenc}
\usepackage[T1]{fontenc}
\usepackage{enumitem}
\usepackage{mathtools}
\usepackage{graphicx}
\usepackage{float}
\usepackage{siunitx}
\usepackage[font={small,it},justification=centering,skip=0pt]{caption}
\usepackage{subcaption}
\usepackage{authblk}
\usepackage{amsmath,amsfonts,amssymb}
\usepackage{algorithm}
\usepackage{algorithmic}
\usepackage{indentfirst}
\usepackage{times} 
\usepackage{geometry}
\usepackage{hyperref}
\usepackage{setspace}
\usepackage{comment}
\hypersetup{
    colorlinks=true, 
    linkcolor=blue, 
    filecolor=magenta, 
   urlcolor=blue,
    citecolor=blue 
}
\begin{document}

\title{Derivative Extrapolation Using Least Squares}
\author{Nick Butler, Ian Faber, Boris Shapoval, Armen Davis, Peter Borrell, Jake Mcgrath}

\maketitle

\section*{Abstract}

Here, we present three methods for differentiating discrete sets from streaming processes, e.g. WIFI. One approach is based on optimization of the well-known Savitzky-Golay algorithm. These methods are tested on synthetic data sets and will be implemented on subsets of a real university campus WIFI data set. The applicability of all methods are discussed, where we provide insights on both some of their benefits and pitfalls. This article ends with our conclusion on which method is better for our WIFI data.

Time series data is generated in practically every discipline, and it comes in countless forms. This means that analyzing time series data is important, and potential new approaches that can be developed and used to extract information from them can lead to recognizing important new patterns and possibly to lead to new insights and discoveries. In this article, novel Least Squares techniques that are "data driven" are presented. The approaches presented provide a powerful framework that allows polynomial regression, and differentiation, of virtually any set of data points, including time series and streaming data. We also present a different point of view on the problem and propose improvements to previous algorithms. To develop the derivative approximations we implement a degree finder that allows us to create low error approximations. Finally, we apply our methods to a real-world WIFI data set and find that our approach based on SSA is the best result.
 
\vspace{0.1in}
\begin{flushleft}
\textbf{Key Words}: Least Squares, Derivative Extrapolation, Savitzky-Golay, Singular Spectrum Analysis (SSA), WIFI data, data-driven methods
\end{flushleft}

\begin{flushleft}
\textbf{Faculty Sponsors University of Colorado Boulder} : Dr. James H. Curry, Departments of Applied Mathematics and Computer Science and Dr. Daniel Massey, Department of Computer Science.
\end{flushleft}

\newpage

\section{Introduction}
Questions of interpolation and differentiation of data are well known. Initially, polynomial interpolations and divided difference methods were considered. Interpolated polynomials must go through all points of the data, and splines, proposed by Schoenberg in 1946, \cite{Shoenburg} were an example of this approach. It was found that for real and "dense data," or data having multiple values given for some points, splines and polynomial fits were messy, hard to calculate, and provided an incorrect pattern. However, Vaughan further explored the method in 1982 \cite{Vaughan} and found that quintic splines are quite good for applications when computing power is limited, yet Vaughan's method was still wasn't ideal. Burden and Faires' numerical analysis textbook \cite{Splines} provides an approachable exploration of the concept of splines in detail in Chapter 3, indicating some of its advantages and pitfalls.

As an alternative to splines, Least Squares proved to be a better method of approximating sections of generic data sets. Savitzky and Golay \cite{Savitzky-Golay} first noticed this possibility in 1964, and their approach led to generating good approximations by splitting up a data curve into numerous windows and calculating a Least Squares regression through each window. They also proposed a method of differentiating data using a table of convoluting integers, described in Appendix I of \cite{Savitzky-Golay}. The Least Squares approach of approximation gained popularity in works such as \cite{SLS}, \cite{Mortari}, \cite{Lower Rank}, \cite{Smoothing}, and \cite{Multi S-G} and is also included in popular text books such as Oliver and Shakiban \cite{Oliver-Shakiban} and Strang \cite{Strang}. Of those references, the most interesting application of Least Squares for this article was presented by Mortari \cite{Mortari} in 2017. Mortari’s application to "unknown differential equations" motivated our thinking about data sets that don’t have known derivatives. While the method proposed by Savitzky and Golay in 1964 is very good at reducing noisy data via Least Squares, computational power and algorithmic strategies have progressed. For example, because of limited computing power, the Savitzky-Golay approach took all points into consideration at the same time. Further, it requires massive tables of integers, which may not be the most intuitive.

The purpose of this article is to describe methods that can determine derivatives of inherently noisy data. For each of the proposed methods, how the noise is reduced and how derivatives are extrapolated will be determined in several different ways and for different orders. This article is structured as follows. Section \ref{PE} will present our first derivative extrapolation technique, Polynomial Extrapolation (PE), which uses Savitzky-Golay as its base. Section \ref{RPE} talks about Reverse Polynomial Extrapolation (RPE), a combination of simple differentiation techniques and the Savitzky-Golay technique. Section \ref{SSA} introduces our third technique of derivative extrapolation, based on Singular Spectrum Analysis (SSA). Section \ref{results} details the results of running each method on different data sets, including two real world WIFI data sets. Finally, Section \ref{discussion} contains our discussion of the methods, their benefits and shortcomings, and possible directions for future research. 

We have not seen these methods in previous literature, and believe they are new. In particular, we believe the degree optimization algorithm we introduce in Section \ref{degree OP method 1} is a new take on Least Squares regression, and our definition and use of window size throughout the article is unique. Also, our approach of using Least Squares as opposed to other numerical derivative methods is a new approach to the problem.

\section{Polynomial Extrapolation} \label{PE}
The most basic method we propose for extrapolating Least Squares derivatives from data, here referred to as Polynomial Extrapolation (PE), is based on a variant of the Savitzky-Golay method's core idea. Savitzky and Golay proposed analyzing a time series by parts, running a polynomial regression of a specified degree on each window slice \cite{Savitzky-Golay}. They identify numerical values for coefficients of desired polynomials by careful tabulation, which are then used in the numerical analysis. We have replaced this with a more modern polynomial differentiation technique. We use a simple formula to find the derivative of a polynomial and approximate the slope of a point in the data, which reduces complexity compared to the giant tables developed by Savitzky and Golay. These polynomial solutions are obtained with a standard Least Squares regression. The parameters needed for each step are the window size and the polynomial degree. These parameters in turn give the rows and columns, respectively, of the matrix used in Least Squares. To describe the window size we use the variable $n$. This variable represents the number of points to the left or right of whatever point we are analyzing. For example, an $n$ value equal to 10 will have 10 points to the left and right of point $i$ resulting in a total window size of 21.
The first step of this algorithm is finding the Least Squares solution of specified window size and degree for point $i$. This solution will be an array as shown below, where $d$ is the degree of the polynomial. 
$$\{c_0,c_1,c_2,...,c_d\}$$
The polynomial function that this describes is
$$f(x) = c_0+c_1x+c_2x^2+...+c_dx^d$$
From here we know by the power rule that the derivative would take the form
$$f'(x) = c_1+2c_2x+...+dc_dx^{d-1}$$
resulting in an estimate for the slope at point $i$ by finding $f'(i)$. This is done for all points $i$ in the algorithm which ultimately generates a curve for the derivative. The pseudocode for this is shown in algorithm \ref{LSDE}.

\begin{algorithm} 
\caption{Least Squares Derivative Extrapolation (X,Y,n,d)}
\begin{algorithmic} \label{LSDE}
\REQUIRE $X$ is an array of domain data, $Y$ is an array of data on that domain, $n$ is an integer number relating to window size, $d$ is degree for polynomial fits.
\FOR{ $i$ $1$ to $size(X)-2n$ }
    \STATE $coef \leftarrow LeastSquares(X[i:i+2n+1],Y[i:i+2n+1],d)$
    \STATE $derivativeValue \leftarrow 0$
    \FOR{ $c$ $1$ to $size(coef)-1$ }
        \STATE $derivativeValue \leftarrow derivativeValue + c*coef[c+1]^{c-1}$
    \ENDFOR
    \STATE $dY[i] \leftarrow derivativeValue$
\ENDFOR
\RETURN $dY$
\end{algorithmic}
\end{algorithm}

While the PE method works with a single specified degree across the entire data set, the "appropriate" degree for each window may differ. For example, a data set composed of curved and linear sections will be better approximated with a mix of degrees, namely linear for the linear sections and higher degree regressions for the curved sections. To find these sets of "ideal" degrees for each window, we developed a degree finder algorithm. Note that for all errors discussed the equation below was used, where $"n"$ is the number of data points and $"f"$ is a predicted polynomial function.

\begin{equation*}
    e = \sqrt{\sum_{i=1}^n \frac{(f(x_i) - y_i)^2}{n-1}}
\end{equation*}

\subsection{Polynomial Extrapolation Degree Optimization} \label{degree OP method 1}
We used Least Squares to find the ideal polynomial degree for each window, based on relative error of one degree to the next. We created this relative error algorithm with two underlying assumptions. The first is that with increasing degrees, the error would always decrease. The second is that the proper degree would be found due to fast convergence of the error versus degree function. We picked the degree, $d$, using algorithm \ref{first degree method}.

\begin{algorithm} 
\caption{Degree Finder(x,y,check)}
\begin{algorithmic} \label{first degree method}
\REQUIRE $x$ is an array of domain data, $y$ is an array of data on that domain, $check$ is an integer number that limits the maximum possible degree.
\STATE $error \leftarrow$ error of linear polynomial fit
\FOR{$d$ from $2$ to $check$}
    \STATE $newError \leftarrow$ error of d-degree polynomial fit
    \IF{$\frac{error}{newError}-1 < \epsilon$}
        \RETURN{$d$-1}
    \ENDIF
    \STATE $error \leftarrow newError$
\ENDFOR
\RETURN $1$
\end{algorithmic}
\end{algorithm}

\begin{figure}[H]
\centerline{\includegraphics[scale=0.4]{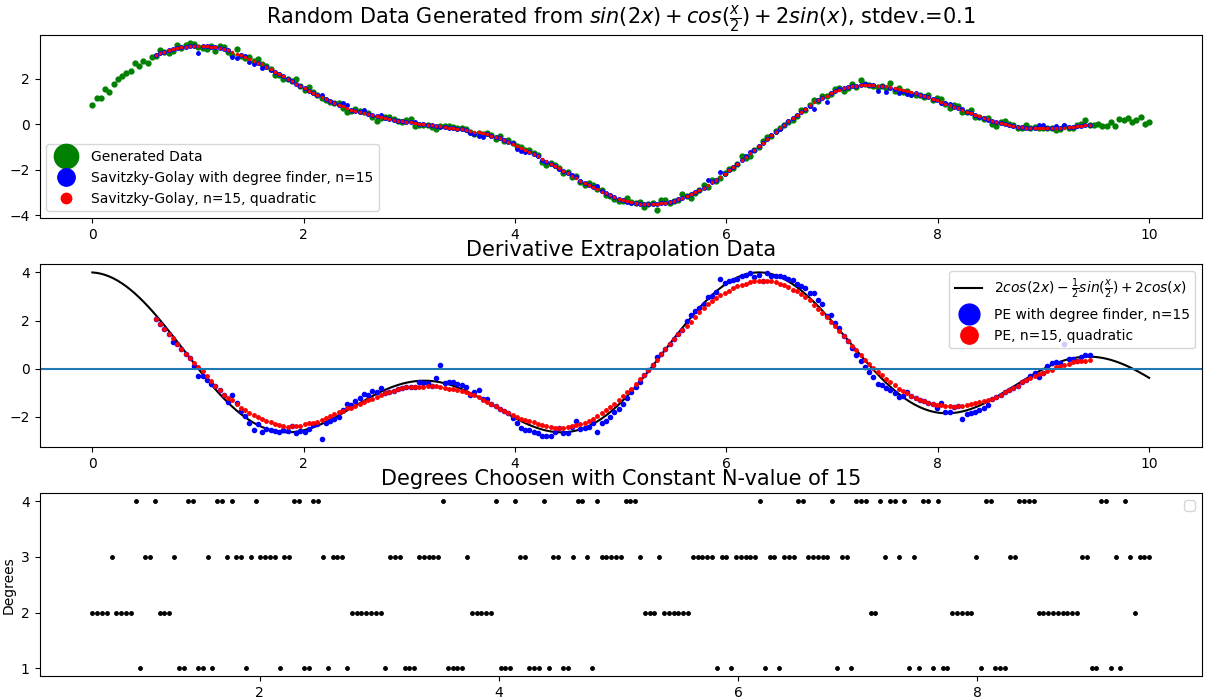}}
\caption{Generated data from a function that shows derivative extrapolation using PE without a degree finder and with a degree finder. The bottom graph shows the degrees chosen when the degree finder was used.}
\label{degree_comp}
\end{figure}

For our numerical experiments, $\epsilon$ was chosen to be $1\%$, meaning that the relative error from the previous degree to the next degree was small enough, namely only a $1\%$ change, that the lower degree was sufficient to describe the data. Figure \ref{degree_comp} shows a comparison between PE with and without the degree finder algorithm, and it can be clearly seen that the degree finder results in a better approximation of the actual derivative curve. However, the degree finder is not as smooth, and may be more susceptible to noise. In terms of the actual degrees picked, we can see that higher degrees perform better when approximating values at the local extrema of the actual derivative. The quadratic PE consistently underestimates the magnitude of the actual derivative at those points, whereas the degree optimized PE better approximates those extrema. However, they seem to do equally well when approximating linear sections of the derivative. The degree finder is not limited to just PE, it can also be applied to the Savitzky-Golay algorithm. This means that we can generate better approximations for both derivatives and data sets themselves.

Our initial assumption that increasing degrees led to decreasing error was found not to always be true. In fact, the error became "chaotic" as higher degrees were chosen. This resulted in no clear trend from degree to degree. This was the strangest result especially when considering the Least Squares application to our problem. Theoretically, when choosing $n+1$ distinct points for an $n$ degree polynomial fit, the error should equal zero. This appeared to rarely be the case in our experiments when $n$ exceeded around 15 or more. However, this potential numerical artifact did not pose a problem for PE, as the degrees chosen were generally smaller than window size.

The second assumption broke down with varying amounts of noise in the data. The noise in the data was determined from a normal distribution to the original curve. When the standard deviation of the data was low, the error versus degree function behaved as expected, where the error converged quickly to zero. On the other hand, when the data had a high standard deviation, the error first decreased slowly, leading to a linear fit most of the time. Though we did not plan for the poor convergence on data with high variance, PE still proved viable as it would not try to over fit these data occurrences with high degree regressions.

\section{Reverse Polynomial Extrapolation} \label{RPE} 
Our Reverse Polynomial Extrapolation (RPE) method is based on reversing the order of PE. In RPE, we first take a simple numerical derivative, then we employ the Savitzky-Golay noise reduction technique. Taking the derivative and removing the noise are two non-commutative operations, making the results of PE and RPE different. The Savitzky-Golay method will remove some random noise that was in the initial data, resulting in a "better final curve." The Savitzky-Golay method can also be used iteratively if further reductions of noise is desired.

\subsection{Reverse Polynomial Extrapolation Mathematics} \label{RPE-math}
The numerical derivative is taken by a simple symmetric quotient
\begin{equation*}
    f'(x_i) \approx \frac{y_{i+1} - y_{i-1}}{x_{i+1} - x_{i-1}}
\end{equation*}

We do not need to know an exact derivative value at every point. Those points are subject to random noise, this makes an exact derivative value far off from the actual function's derivative value. A major concern is to have derivative values be accurate with a "similar error," and the symmetric quotient rule satisfies that requirement.

We expect that derivative data will be inaccurate due to the noise in the original data. However, that chaotic behaviour will be removed with our "best variation" of the Savitzky-Golay algorithm discussed above, i.e. the algorithm with the degree finder. It will remove the noise and in the end we will have an accurate derivative trend.

\subsection{Number of noise reductions}
When taking the derivative of noisy data, the derivative will have more noise than the data set itself, and that is exactly what RPE does. To fix this, we introduced more iterations of Savitzky-Golay method to make sure that our result is actually smooth and uninfluenced by noise from differentiation. Choosing the correct number of extra iterations isn't trivial, because after some iteration reducing the noise will only lock the curve on a fake trend.

Since PE removes noise before differentiating, there is no extra noise introduced once the derivative is actually taken. However, since RPE differentiates first, more noise is introduced and it will always be noisier than PE. To fix this and get similar behaviour from RPE, we applied the Savitzky-Golay method twice after differentiating. The first iteration removes the extra noise that came from differentiation, and the second iteration ensures that RPE and PE are based on similarly noisy data.

\section{Derivative Extrapolation Using Singular Spectrum Analysis} \label{SSA}

Singular Spectrum Analysis (SSA) is helpful for time series data, as it can split that data into components. To split the data into components, SSA utilizes Singular Value Decomposition (SVD). For a much more in-depth dive into all things SSA, see \cite{SSA ref}. We thought to use this method in derivative extrapolation because of its ability to reduce Gaussian noise from data. The first step of SSA is to put time series data into a Hankel data matrix, which is then the input of an SVD. For our application, the number of columns of this Hankel matrix represents the window size. Below shows how a time series array, $X$, is converted to the Hankel data matrix, $H$.

$$X = [x_1,x_2,...,x_{end}] \to H_{n \times m} = 
\begin{bmatrix}
x_1&x_2 &\hdots &x_m\\
x_2&x_3 &\hdots &x_{m+1}\\
x_3&x_4 &\hdots &x_{m+2}\\
\vdots&\vdots &\ddots &\vdots\\
x_n&x_n+1 &\hdots &x_{end}\\
\end{bmatrix}$$

The next step in our process is removing components that contribute mostly Gaussian noise. Each component has an eigenvalue associated with it. The larger the eigenvalue, the more the component explains the data. The smaller the eigenvalue, the more it contributes to noise. If all components are used, we would say that 100\% of the variance is explained. To help remove noise, we only use the components that contribute to between 90\% and 99\% of the variance. The results in this article commonly used a "variance explained" value of between $0.95$ or $0.975$.

Once the standard steps of SSA have been performed and most of the Gaussian noise is removed, we can find the derivative. From SSA we have a constant number of components, where each row in a component represents a window surrounding the point where we want to find a slope. Typically at this point in the process of SSA, the components are converted back to a time series by averaging diagonal cells of the Hankel matrices. By averaging these cells, there is a loss of information. Our derivative extrapolation method attempts to preserve this information by skipping the process of averaging the diagonals. Instead, our algorithm will loop over each component, extract a row, perform a Least Squares linear regression to find a slope, and sum these slopes for each component. For example, if we desire to estimate the derivative at point $i$, we perform a linear regression on the $i$th row of each component and sum the results.

\begin{algorithm}
\caption{SSA derivative(X,Y,n,var=0.95)}
\begin{algorithmic}
\REQUIRE $X$ and $Y$ are arrays holding the data, $n$ is an integer number relating to window size, $var$ is a floating point number between 0 and 1 giving variance explained.

\STATE $window \leftarrow 2n+1$
\STATE $H \leftarrow$ hankel matrix using array $Y$ with number of columns equal to $window$
\STATE $U,E,V \leftarrow$ SVD on H
\STATE $numEigenvalues \leftarrow$ number required for $var$ variance to be explained
\STATE $Components \leftarrow$ array of matrices of length $numEigenvalues$ \\ \quad where
        a matrix is equal to $E[idx]*(U_{:,idx}\times V_{idx,:})$

\FOR{ $i$ $1$ to $numEigenvalues$ }
        \FOR { $j$ $n$ to $length(X) - n$}
                \STATE $slope \leftarrow LeastSquaresLinearFit(X[j-n][j+n],Components[i][j])$
                \STATE $dY[j] += slope$
        \ENDFOR
\ENDFOR
\STATE $dY \leftarrow dY$ shifted due to the hankel matrix
\RETURN $dY$
\end{algorithmic}
\end{algorithm}

\section{Results} \label{results}

\begin{figure}[H]
\centerline{\includegraphics[scale=0.4]{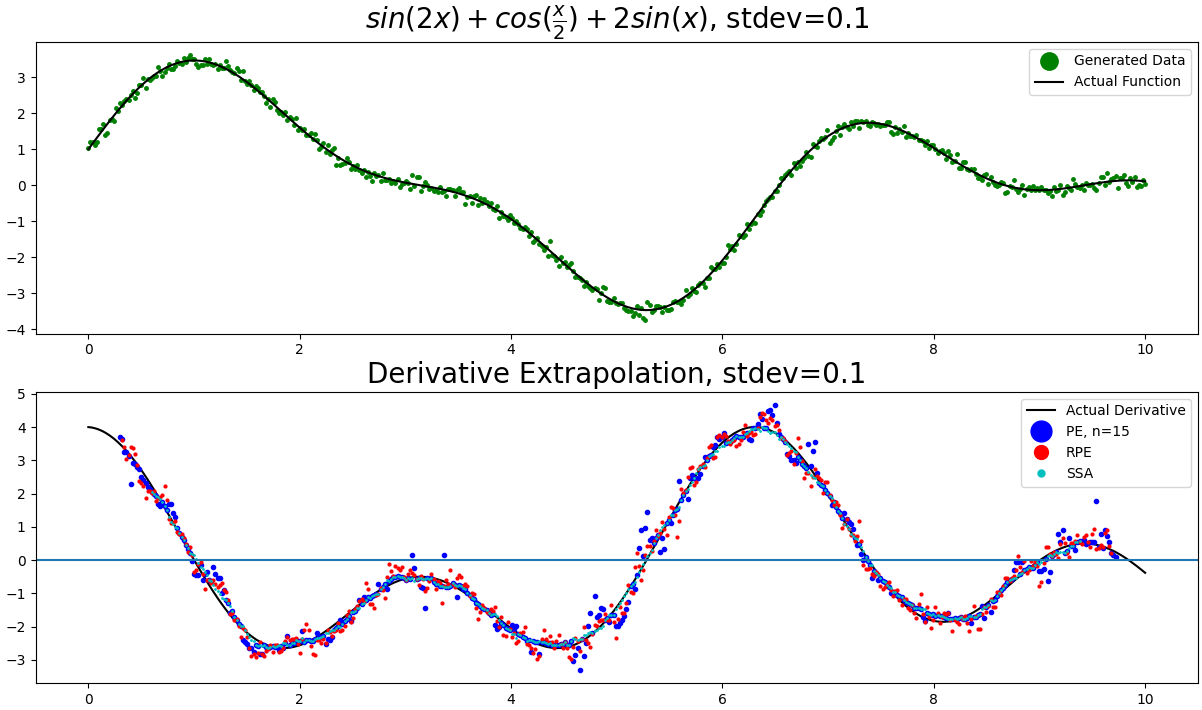}}
\caption{Top: A graph showing the original function, which consists of multiple components, that we want to find the derivative for. Bottom: A graph showing the derivative extrapolation of the test function with the three proposed methods. The noise in the top graph is based on a standard deviation of 0.1.}
\label{r1}
\end{figure}

In Figure \ref{r1}, we use test data with 3 main components which are in the form of sine or cosine functions, namely, $sin(2x)$, $cos(\frac{x}{2})$, and $2sin(x)$. PE and RPE seem to straddle the actual derivative equally above and below, showing no apparent systematic error. For SSA, the bottom graph shows an almost perfect fit of the line with minimal amounts of noise. We can also see the effect of window size in figure \ref{r1} in the form of a "delay" at each end of the derivative graph. Since the window size we've chosen is 15 points, we can see a corresponding delay of 7 points on the left and right sides of the generated derivative curve. This means that in order to capture derivative information at the extremes of a data set, either a smaller window size can be chosen or the domain of the original data curve can be increased. Regardless, there will always be a delay with these methods.

\begin{figure}[H]
\centerline{\includegraphics[scale=0.4]{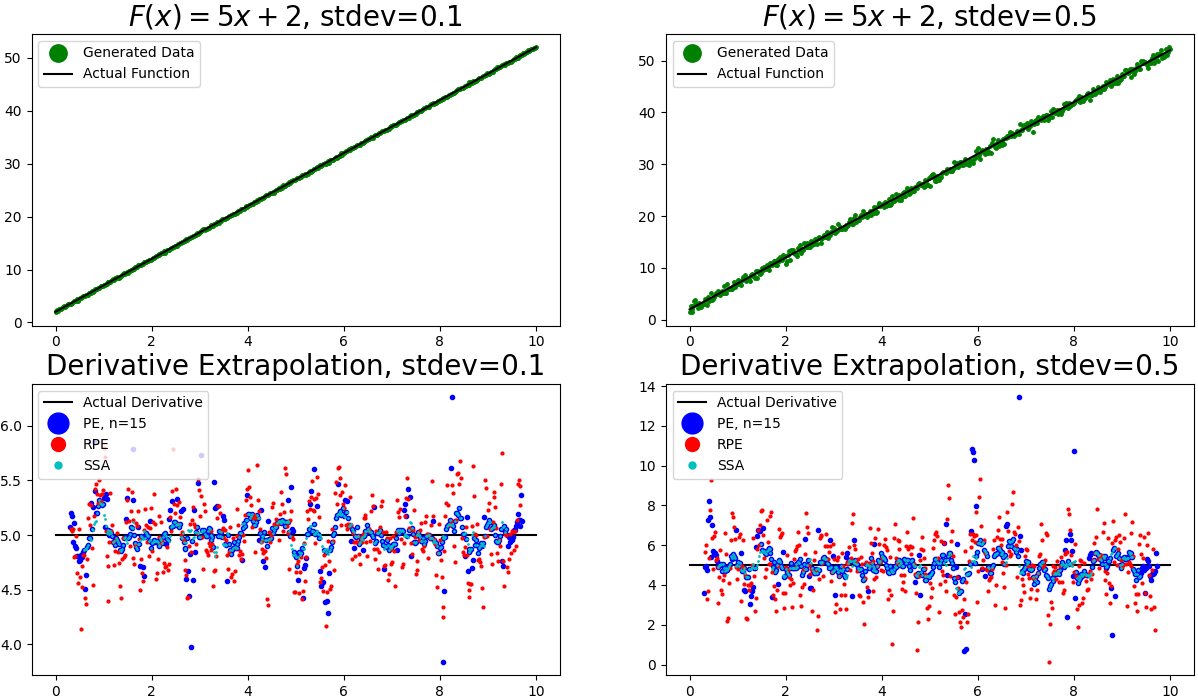}}
\caption{Top: Two graphs showing a linear function we want to find the derivative for. Bottom: Two graphs showing the derivative extrapolation of the test function with the three proposed methods. The left and right graphs show different levels of noise.}
\label{r2}
\end{figure}

The trends we see in this case may not agree with real life, due to the random nature of the error in this case and more systematic issues in real life. Figure \ref{r1} only gives us a general feeling of the performance of the methods. In real life we would also expect longer linear parts and quicker falls and rises, which our test case doesn't have. No matter what we do, a fake data set will never prepare us for real world features. Nevertheless, we will look upon another fake data set to get a feeling of how well different methods estimate linear trends. To see if small oscillations are inherent to the sinusoidal curve or the noise in any data, we performed these methods on a simple linear line shown in Figure \ref{r2}.

In Figure \ref{r2}, we see that there are oscillations in the data from all explored methods. This means that small oscillations in the derivative extrapolation must be attributed to the noise in the data. For larger amounts of noise in the data, we see that there appears to be increased amounts of oscillations above and below the actual solution curve. This is expected, as the methods are following less clear data trends when the data curve is noisy. We can also see that in the case of the curve with more noise (larger standard deviation), PE does much better than RPE, while PE and SSA seem to follow similar trends.  We can also see that the magnitude of oscillations increases as the ratio of standard deviations increases. In this case, the ratio of standard deviations is 5, and we see a corresponding increase in magnitude by about the same ratio. When comparing the three methods here, SSA shows the greatest reduction in noise. This makes sense, as the main goal of SSA is to take out Gaussian noise from time series data. The PE method does fairly well and "matches" SSA in many places; however, it tends to have large jumps away from the true solution since it does not pass over the noise in the data well.
\vspace{5mm}

\begin{figure}[H]
\centerline{\includegraphics[scale=0.35]{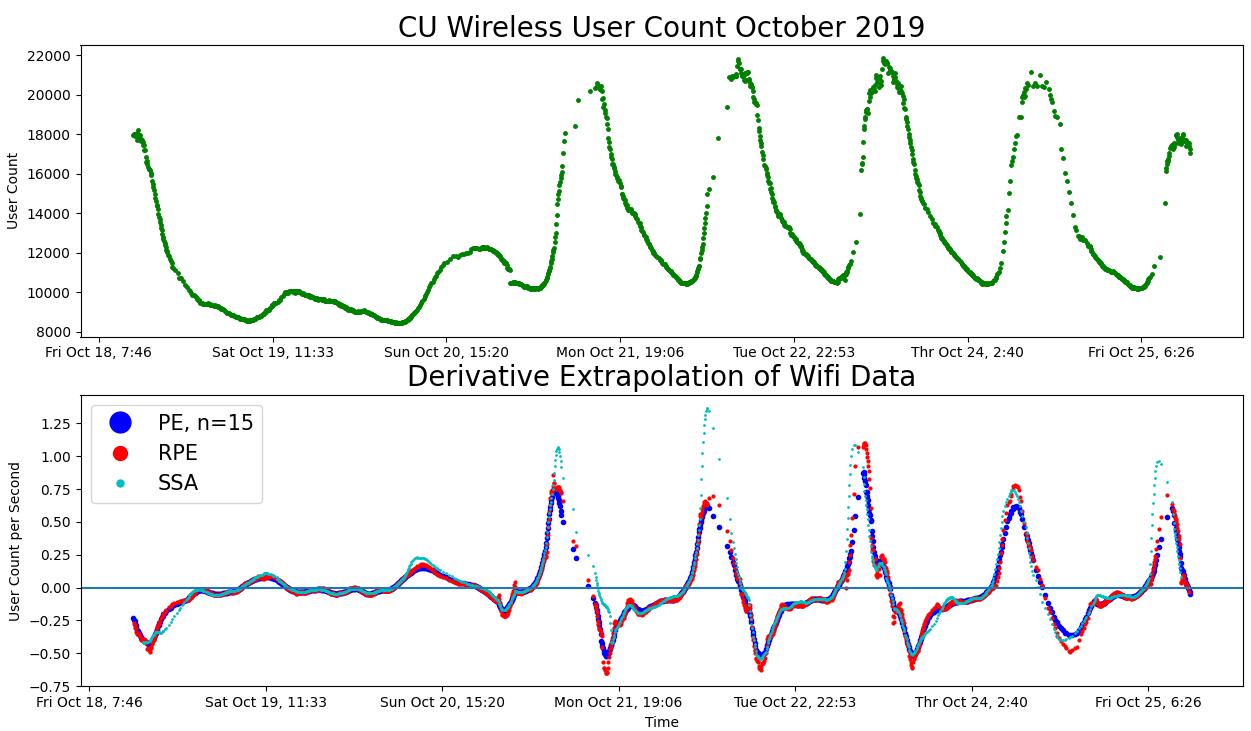}}
\caption{Top: Data of the total Wi-Fi user count on the CU Boulder wireless network. Bottom: Derivative extrapolation using three methods on a real data set with a small amount of noise.}
\label{r3}
\end{figure}

\begin{figure}[H]
\centerline{\includegraphics[scale=0.35]{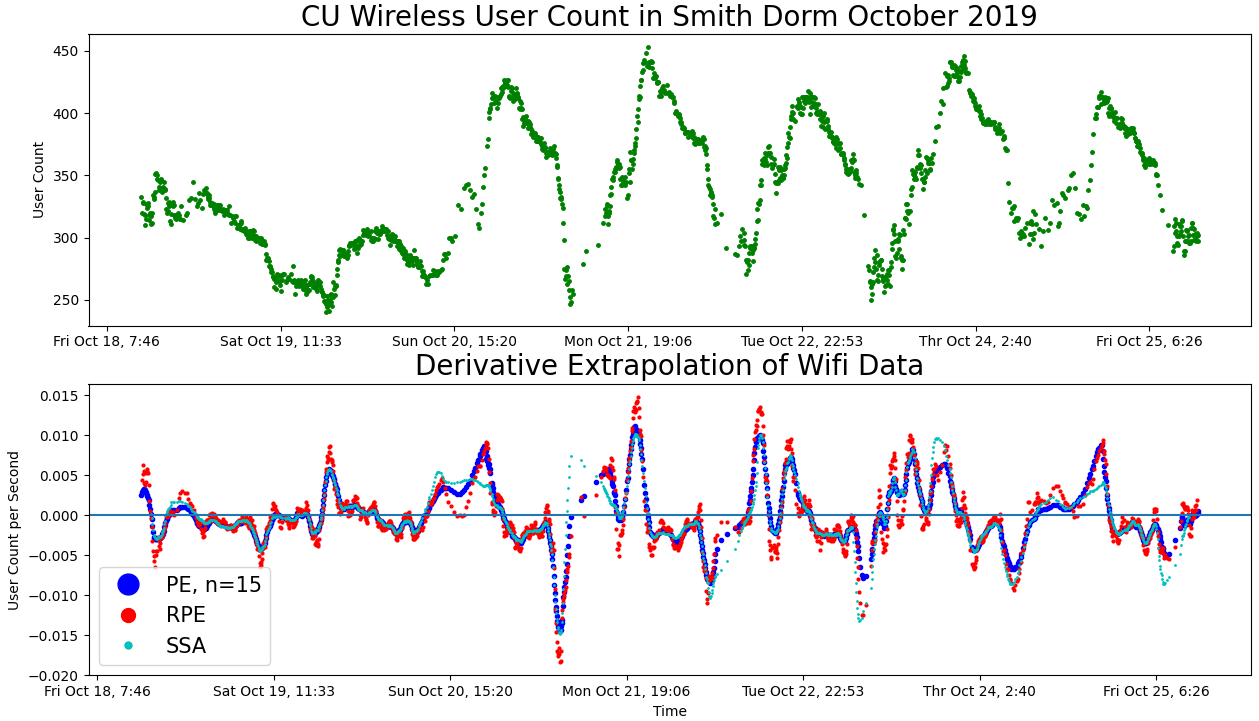}}
\caption{Top: Data of user count on CU wireless network in Smith Dorm. Bottom: Derivative extrapolation using three methods on a real data set with a large amount of noise.}
\label{r4}
\end{figure}

After verifying our methods with the test data sets and noting their advantages and disadvantages, we moved onto our "real" data set of Wi-Fi usage at the CU Boulder campus. The campus has three different networks: CU Wireless, CU Wireless Guest, and Eduroam. CU Wireless is the most used network, and we focused on it exclusively in our analysis. The data was collected via routers distributed over the campus territory and buildings. A number of connected users is measured at times with no particular periodicity and is summed up over all routers to acquire a total number of users across the campus, or in a particular building. We analyzed both the full network and an average dorm, Smith Dorm \cite{Wi-Fi}, during a full week in October of 2019. The real world data used had personal identification characteristics removed prior to processing the data. The information used has been stripped of any information other than what is required for counting active users, a general datum of where the count was taken, and the time when the value was obtained. It is not possible to identify a movement of any of the real data recorded since there is equal likelihood of them staying, or moving to another location measured. This data set has no known derivative, and we will use our methods to determine mathematically reasonable approximations. The full network data set is shown in the top half of Figure \ref{r3}, and the Smith Dorm data set is shown in the top half of Figure \ref{r4}.

Figure \ref{r3} shows a smoother curve and Figure \ref{r4} shows a more "jittery" and noisy curve. However, the curves show similar trends based on the day of the week. We can quickly see that in both cases, RPE is the most noisy derivative extrapolation method. The RPE curve tends to mimic the PE approximation the most; however, RPE is better at showing peaks in the derivative. On the whole, SSA seems to be much smoother than both PE and RPE, but it differs significantly from both in multiple places. 

In the case of overall network trends from Figure \ref{r3}, we can see that our methods create trends that match what we'd expect for a typical week at a university, namely, more rapid changes in network traffic during the week and significantly less on the weekends. We can see a clear increase in the number of users per second after the weekend, which can be explained by students coming back to campus after taking the weekend off. More students on campus equates to more devices connecting to the network, which our methods correctly show as a higher magnitude of users per second. Likewise, when students leave campus for the day the number of devices connecting to the network per second becomes negative, which we'd expect with so many devices disconnecting from the network. Our methods also correctly show critical points where we'd expect, namely the "humps" of each day.

In terms of network trends in the dorm from Figure \ref{r4}, despite the messiness of the curves we can still see patterns that we would expect for a typical dorm, i.e. more heavy network use in the evenings and early morning hours than during class times. On the weekends, particularly Fridays and Saturdays, the oscillations settle to small values, meaning that not too many people are leaving the dorms if they've decided to stay in them over the weekend. Understandably, many students choose Sunday to catch up on work and/or get ahead for the week, which is accordingly reflected by higher network traffic and users per second. When students leave for class, we can correctly see a negative spike in users per second, which means devices are disconnecting from Smith's section of the network. Likewise, when they come back after class we can see corresponding spikes in positive users per second.

\section{Discussion} \label{discussion}
\subsection{Comparison of the Three Methods}
In carrying out our experiments, we found that the best of our methods for differentiating a curve via Least Squares was SSA. In general, the PE and SSA methods are most closely related to each other, as for many parts of both curves they are nearly identical. Where they differ is when the data becomes messy. In Figure \ref{r3}, the messiest parts of the data are when there are positive slopes and local maxima. Though PE does well to approximate these sections, it does not show a uniform and smooth derivative. This is easily seen on the sections that the SSA method is larger than PE. PE flattens out here whereas SSA shows clear sinusoidal behavior. Figures \ref{r1} and \ref{r2} also show the effectiveness of SSA in approximating a well-known derivative. Whereas PE and RPE oscillate wildly around the known derivative, SSA matches the derivative incredibly closely in context of both large and small noise standard deviations. Similar to other diagrams, in Figure \ref{r4}, there is a clear distinction between SSA and PE on noisy parts of the data. We see that on sharp, "almost" non-differentiable parts of the data, PE tends to show rounded off parts whereas SSA appears to accurately show the proper peaks and troughs of the cyclical data set. RPE has the most amount of noise and overestimates the derivative's peaks, which is an indication that this method is the worse one.

We believe that the SSA-based strategy is the best method because it exhibits better long-term behavior, continuity, and noise reduction than RPE or PE. Long-term behavior in this case refers to the algorithm's ability to correctly calculate the derivative in context of points outside its window size. This inherently makes sense when looking at how each method works. For RPE, it only uses one point to either side and relies on Savitzky-Golay to take into account a large window size. PE is slightly better than RPE since it initially looks at a larger window size when approximating the derivative, yet it is still limited to that window. Lastly, SSA does best since its uses the data as a whole and can best recognize long-term behavior. The most significant problem with RPE is that it struggles to reduce noise and it therefore tends to over-fit the data. The main problem with PE is it lowers magnitudes at peaks and troughs. PE does not find large slopes as well as SSA and tends to dull parts of the curve as a result.

\subsection{Other Methods}
There are some other strategies that we consider valuable, but did not look at in this article. One is an improvement of the degree finder in the way that it will favor higher degrees by adding some type of weights to the system. The weights would ideally avoid most of the small perturbations that caused the oscillations seen in Figure \ref{r2} and approximate long linear parts better. Another way to remove extra zeros calculated from noise, and hence reduce oscillations, is to implement a minimum - maximum - saddle point finder and make the derivative zero at those points. Then, we will only have one unknown coefficient to build the derivative as a polynomial from its zeros. That last coefficient can be found, again, with linear regression techniques by solving $y_i = b*(x_i-r_1)...(x_i-r_r) = b\Bar{x}_i$ for $b$, given $r_j$ to be the roots and $y_i$ random derivative samples measured with a symmetric quotient rule. We believe that both of these methods may be valuable alternatives to PE and RPE methods in the long term behaviour, but their behaviour in context of quick changes is yet to be discovered and would be a beneficial area of future investigation. We also thought about a potential method that combines SSA and PE. Currently, we only employ a linear regression on each component of SSA decomposition, but it may be beneficial to optimize that degree of fit using PE's degree finder.

When we used the SSA method to find derivatives, we compared it to using Non-Negative Matrix Factorization (NMF)\cite{NMF}. Not surprisingly, these two methods were very similar to each other. The reason we used SSA instead of NMF is for two reasons. First, NMF requires many iterations to converge to a solution; this makes the computation time of NMF much longer than SSA. Second, by reducing noise similar to our SSA method, NMF will consistently underestimate the real derivative due to its requirement of only non-negative values. Hpwever, further investigation may be beneficial.

\section{Conclusion}
In this article, we developed three new methods for calculating derivatives of data sets: Polynomial Extrapolation, Reverse Polynomial Extrapolation, and Derivative Extrapolation using Singular Spectrum Analysis. All methods incorporate the technique of Least Squares applied to streaming data. The PE method first uses Savitzky-Golay to reduce the noise of data, then finds the derivative using a Least Squares solution of linear degree polynomial or higher. In developing PE, we also found a way to optimize the degree of fit for the Least Squares problem by comparing relative error from one fit to the next. We have not seen this degree-based approach in previous literature, as well as applying a sliding window to the differentiation process. We are confident that both approaches are new. RPE first finds the "symmetric derivative," then uses the Savitzky-Golay method twice to reduce noise. Lastly, SSA was used to take into account the whole data set and find linear fits for each component of the data. After developing these methods, we applied them to a real-world Wi-Fi data set and found trends in network activity that correspond to real-world activity. In addition, the methods provided some insights into the structure of the CU Boulder campus' inner life. We concluded that the SSA derivative method was best since it reduced the largest amount of Gaussian noise and showed better long-term behavior.

\section{Acknowledgement}

The research reported here was supported in part by a contract from Sandia National Laboratory to the University of Colorado Boulder, Departments of Applied Mathematics and Computer Science. Special thanks to our colleagues in the University of Colorado Office of Information Technology, Orrie Gartner and Glenn Rodrigues, who made the anonmyized data presented in figures \ref{r3} and \ref{r4} available and helped with many discussion and guidance on data policies and the broader research effort of which this is a part. Special thanks to Maxim Moghadam for making his undergraduate Computer Science thesis material available to the research team.


\begin{thebibliography}{}

\bibitem{Splines} Burden, R. L., \& Faires, J. D. (2011). Interpolation and Polynomial Approximation. In Numerical Analysis (pp. 105–173). essay, Thomson Brooks/Cole Cengage Learning. 

\bibitem{Lower Rank} Gabriel, K. Ruben, and S. Zamir. “Lower Rank Approximation of Matrices by Least Squares with Any Choice of Weights.” Technometrics, vol. 21, no. 4, 1979, pp. 489–498. JSTOR, www.jstor.org/stable/1268288. Accessed 28 July 2021.

\bibitem{SSA ref} Golyandina, N., Nekrutkin, V. V., \&; Zhigljavsky, A. (2001). Analysis of time series structure: SSA and related techniques. Chapman \& Hall/CRC. 

\bibitem {SLS} Lawson, Charles L. and Hanson, Richard J. (1995). Solving Least Squares Problems (book) \&
Publisher, Society for Industrial and Applied Mathematics.

Eprint, 
https://epubs.siam.org/doi/pdf/10.1137/1.9781611971217

Url
https://epubs.siam.org/doi/abs/10.1137/1.9781611971217

\bibitem{NMF} Lee, D.D. and Seung, H.S. (1999) Learning the Parts of Objects by Nonnegative matrix Factorization. Nature, 401, 788-791.

\bibitem{Wi-Fi} Mcgrath, J., Davis, A., Curry, J., Gartner, O., Rodrigues, G., Spielman, S., and Massey, D., "(\textbf{In preparation}) A Wi-Fi Ecosystem: IoT Device Count, Mobile Weather, Clusters of Building: University of Colorado Boulder - Fall Semester 2019 to Spring Semester 2020 - Including a Case Study of the Campus Response to the COVID-19 Perturbation," 2021, manuscript.

\bibitem{Mortari} Mortari, D. “Least-Squares Solution of Linear Differential Equations.” Mathematics 5.4 (2017): 48. Crossref. Web.

\bibitem{Oliver-Shakiban} Olver, P, and Shakiban, C. "Applied Linear Algebra, Second Edition." Springer, 2019, ISBN-13: 978-3319910406, ISBN-10: 331991040X 

\bibitem{Smoothing} Press, W. H., \&; Teukolsky, S. A. (1990). Savitzky-Golay smoothing filters. Computers in Physics, 4(6), 669. https://doi.org/10.1063/1.4822961 

\bibitem{Savitzky-Golay} Savitzky, A., and Golay, M.J.E. “Smoothing and Differentiation of Data by Simplified Least Squares Procedures.” Northwestern University, The Perkin-Elmer Corporation, 1964, xrm.phys.northwestern.edu/research/pdf\_papers/1964/savitzky\_analchem\_1964.pdf.

\bibitem{Shoenburg} Schoenberg, I.J. (1988) Contributions to the Problem of Approximation of Equidistant Data by Analytic Functions. In: de Boor C. (eds) I. J. Schoenberg Selected Papers. Contemporary Mathematicians. Birkhäuser, Boston, MA. https://doi.org/10.1007/978-1-4899-0433-1\_1

\bibitem{Multi S-G} Shekhar, C., “On simplified application of multidimensional Savitzky-Golay filters and differentiators,” AIP Conf. Proc. 1705, 020014 (2016). https://doi.org/10.1063/1.4940262,

\bibitem{Strang} Strang, G. "Linear Algebra and Learning from Data, First Edition." Gilbert Strang, ISBN-13: 978-0692196380, ISBN-10: 0692196382 

\bibitem{Vaughan} Vaughan, C L. “Smoothing and differentiation of displacement-time data: an application of splines and digital filtering.” International journal of bio-medical computing vol. 13,5 (1982): 375-86. doi:10.1016/0020-7101(82)90003-4

\end{thebibliography}
\end{document}